\documentclass[reprint,prx,aps,amsmath,amssymb,floatfix,superscriptaddress,longbibliography]{revtex4-2}
\usepackage{dcolumn}
\usepackage{bm}
\usepackage{graphicx}
\usepackage{amsthm}
\usepackage{color}
\usepackage{xcolor}
\usepackage{verbatim}
\usepackage{esint}
\usepackage[english]{babel}
\usepackage{amsfonts}
\usepackage{slashed}
\usepackage{latexsym}
\usepackage{bm}
\usepackage[colorlinks = true,
            linkcolor = red,
            urlcolor  = blue,
            citecolor = magenta,
            anchorcolor = red]{hyperref}
\usepackage{esint}
\usepackage{soul}
\usepackage{cancel}
\usepackage[normalem]{ulem}
\usepackage{empheq}
\usepackage{dsfont}
\usepackage{amsmath}
\usepackage{ulem}
\usepackage{epsfig}
\usepackage{enumerate}%\usepackage{eulervm} 
\definecolor{sapphire}{rgb}{0.03, 0.03, 0.41}

\DeclareMathAlphabet\mathbfcal{OMS}{cmsy}{b}{n}
\def\be{\begin{equation}}
\def\ee{\end{equation}}

\begin{document}

\title{Continuous Measurement Boosted Adiabatic Quantum Thermal Machines}
% Force line breaks with \\
%\thanks{notes}%

\author{Bibek Bhandari}
\affiliation{Department of Physics and Astronomy, University of Rochester, Rochester, NY 14627, USA}
\affiliation{Institute for Quantum Studies, Chapman University, Orange, CA 92866, USA}
\author{Andrew N. Jordan}
\affiliation{Institute for Quantum Studies, Chapman University, Orange, CA 92866, USA}
\affiliation{Department of Physics and Astronomy, University of Rochester, Rochester, NY 14627, USA}

%\altaffiliation[Also at ]{Physics Department, XYZ University.}%Lines break automatically or can be forced with \\

% \email{Second.Author@institution.edu}
%\affiliation{%
% Authors' institution and/or address\\
% This line break forced with \textbackslash\textbackslash}%

 %\homepage{http://www.Second.institution.edu/~Charlie.Author}
%\affiliation{Second institution and/or address\\
% This line break forced% with \\}%
%\affiliation{
% Third institution, the second for Charlie Author
%}%

%\date{\today}% It is always \today, today,
             %  but any date may be explicitly specified
             
\begin{abstract}
We present a unified approach to study continuous measurement based quantum thermal machines in static as well as adiabatically driven systems. We investigate both steady state and transient dynamics for the time-independent case. In the adiabatically driven case, we show how measurement based thermodynamic quantities can be attributed geometric characteristics. We also provide the appropriate definition for heat transfer and dissipation owing to continous measurement in the presence and absence of adiabatic driving. We illustrate the aforementioned ideas and study the phenomena of refrigeration in two different paradigmatic examples: a coupled quantum dot and a coupled qubit system, both undergoing continuous measurement and slow driving.  In the time-independent case,  we show that quantum coherence can improve the cooling power of measurement based quantum refrigerators. Exclusively for the case of coupled qubits, we consider linear as well as non-linear system-bath couplings. We observe that non-linear coupling produces cooling effects in certain regime where otherwise heating is expected.  In the adiabatically driven case, we observe that quantum measurement can provide significant boost to the power of adiabatic quantum refrigerators. We also observe that the obtained boost can be larger than the sum of power due to individual effects. The measurement based refrigerators can have similar or better coefficient of performance (COP) in the driven case compared to the static one in the regime where heat extraction is maximum.  Our results have potential significance for future application in devices ranging from measurement based quantum thermal machines to refrigeration in quantum processing networks. 
\end{abstract}

%\pacs{Valid PACS appear here}% PACS, the Physics and Astronomy
                             % Classification Scheme.
%\keywords{Suggested keywords}%Use showkeys class option if keyword
                              %display desired
\maketitle

\section{Introduction}
\label{sec:intro}
Thermal transport and heat-to-work conversion in quantum systems has been extensively studied theoretically as well as experimentally owing to its significance in various fields ranging from quantum thermal machines\cite{scovil1959,geusic1967,pendry1983,
kieu2004,
bibekminimal,benenti2017,
alicki1979,schwab2000,meschke2006,blickle2012,
brantut2013,martinez2016,
rossnagel2016,thierschmann2015,josefsson2018} to quantum information processing\cite{partanen2016,tan2017}. Likewise, continuous quantum measurement has also been a modern subject of interest, investigated both theoretically\cite{wiseman2009,barchielli2009,jacobs2014,gisin1984,diosi1988,
wiseman1996,chantasri2013,chantasri2015} and experimentally\cite{gisin1993,gambetta2008,murch2013,weber2014,hacohen2016}, disclosing applications to quantum control\cite{taylor2017,hacohen2018,minev2019} as well as thermal transport in nanoscale devices\cite{jacobs2009,yi2017,cyril2018,ding2018,solfanelli2019,das2019,
debarba2019,buffoni2019,seah2020,bresque2021,sreenath2021}. The effect of continuous quantum measurements on thermal transport and heat-to-work conversion in driven as well as static systems will be the primary focus of this paper.  

Measurement based thermal machines in itself is a recently introduced subject. However, most of the works are based on strong projective measurements\cite{chand2017,chand2018,cyril2018,solfanelli2019,das2019,debarba2019,buffoni2019,
seah2020,bresque2021}. Some research have also been done to study the steady state devices based on continuous measurement\cite{yi2017,sreenath2021,hasegawa2020,cyrilqm}. Connecting a monitor to the system, one can make weak generalised measurements revealing some but not complete information about the system. Undergoing continuous quantum measurement changes the state of the sytem and hence the dynamics of the device. One can utilize this unique property of quantum measurement, monitoring an observable incompatible with energy, to alter the energy transport in nanoscale devices.

In the framework of open quantum systems, adiabatic thermal machines can be obtained by slow modulation of device parameters, where the time scale associated with driving is larger than the relaxation time of the system\cite{cavina2017,brandner2020,bibekgeo,bibekgreen}. It has been observed that adiabatic driving can be utilized to pump heat from one bath to another, obtaining heat engines and refrigerators\cite{janinedot,hanggi}. Moreover, adiabatically pumped heat current per cycle is geometric in nature and does not depend on the the details of the driving\cite{bibekgeo,janinedot}. However, the magnitude of heat extraction is rather small due to the slow nature of driving. Note that the performance of adiabatic quantum thermal machines can be improved by using shortcuts to adiabaticity techniques\cite{shortcut1,shortcut2}.

\begin{figure}[t]
\includegraphics[width=1\columnwidth]{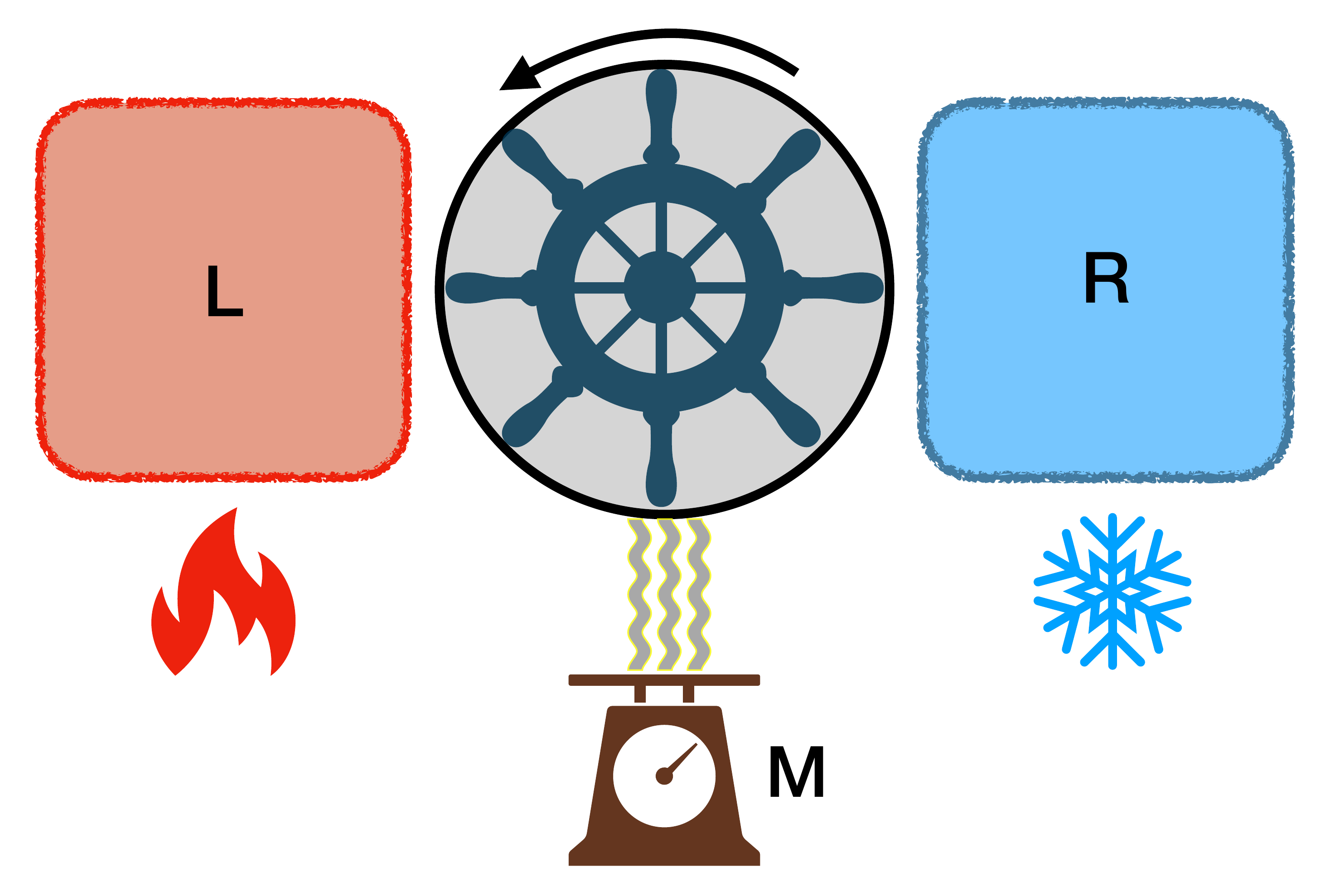}	
	\caption{An adiabatically driven system attached to two thermal baths; the bath on the left (L) hand side is kept at temperature $T_{\rm L}$ and on the right (R) hand side is kept at temperature $T_{\rm R}$. We will consider, $T_L\geq T_R$. In the meantime, the system is being continuously monitored by a measurement apparatus (M).}
	\label{fig:setup}
\end{figure}

In this paper, we present a comprehensive study of whether the phenomena of measurement can be utilized to enhance the performance of adiabatic quantum thermal machines or to even realize a quantum thermal machine per se.  Under the Born-Markov approximation, the contributions from the thermal baths and the measurement probe to the master equation are additive\cite{jacobs2014}. We expand the master equation at different orders of driving frequency to obtain the contribution from adiabatic driving\cite{bibekgreen}. We present a theoretical formulation that unifies two different means of powering a thermal machine, namely quantum measurement and adiabatic driving. We also show how measurement based thermodynamic quantities can be attributed geometric characteristics in the adiabatic limit. Our work is important to understand and control thermal transfer in slowly-driven quantum circuits as well as to improve the power and performance of measurement based thermal machines. Considering two different setups based on coupled quantum dots and couped qubits, we show that continuous quantum measurement can not only significantly boost the power of adiabatic quantum refrigerators but also improve the coefficient of performance under suitable conditions. It has been previously observed that quantum coherence can improve the power of refrigeration only in highly non-equilibrium conditions\cite{uzdin2015,brandner2017}. We observe that under
continuous monitoring, quantum coherence can help improve the power of refrigeration even when the baths are maintained at same temperature.

%We study both time-independent and adiabatically driven case. In the time-independent  case we study both steady state and the transient regime. While considering the steady state, we discard the measurement record.  For the coupled qubit setup, we consider linear as well as non-linear coupling Hamiltonian. We observe that with non-linear coupling, one can obtain cooling in certain regime where otherwise heating would be expected. In the set up based on quantum dots, we observe that quantum coherence enhances the cooling effect. In the adiabatically driven case, we consider only the steady state and discard the measurement record. 

This paper has been organised in the following order. In the next section, we will present the model under investigation.  In Sec.~III, we will study the combined effect of adiabatic driving and continous measurement in open quantum systems.  In Sec.~IV, we will define different thermodynamic quantities relevant to the setup considered. In Sec.~IVA, we will study the geometric nature of work done by continuous measurement in the presence of adiabatic driving. In Sec.~V, we will utilize the formulation developed in the previous sections for two different paradigmatic examples, namely a coupled quantum dot system attached to fermionic baths and coupled qubit system attached to bosonic baths. In Sec.~VI, we will present the results for the coupled quantum dot system. We will study both time-independent and driven cases. For the time-independent case, we will study the transient dynamics and the effect of coherence on the measurement induced refrigeration. In Sec.~VII, we will present some results for the coupled qubit setup. The conclusions will be drawn in Sec.~VIII.
\section{Model}
\label{sec:model}
The setup we consider is sketched in Fig.~(\ref{fig:setup}). The central system consists of a quantum system whose energy eigenstates are represented by the velocity vector $\vec{U}(t)$. The velocity vector is driven adiabatically, such that the driving frequency is associated with the smallest energy scale pertaining to the setup.
We consider that the quantum system is in contact with two thermal baths.  The Hamiltonian for the setup without the measurement probe is given by
\begin{equation}
{ H}(\tau)= H_S(\tau)+H_B+H_C.
\end{equation}
where the system Hamiltonian is
\begin{equation}
H_S(\tau)=\vec{F}\cdot \vec{U}(\tau),
\end{equation}
where $\vec{F}$ is the force vector which can be written in terms of projection operators onto the eigen basis of the system Hamiltonian\cite{bibekgeo,janinedot}. The bath Hamiltonian $H_B=H_{B,L}+H_{B,R}$ consists of two terms corresponding to the two baths $L$ and $R$.  The baths are considered to have continuous degrees of freedom and a finite temperature ($T_L$ for the left bath and $T_R$ for the right bath). We consider, $T_R\leq T_L$. The baths can be of bosonic or fermionic nature: in the later case, the chemical potential will be fixed to zero, $\mu_L=\mu_R=0$.  The Hamiltonians of the bosonic and the fermionic baths are given by
\begin{align}
H_B^{(F)}&=\sum_{k,\alpha=L,R}\epsilon_{k\alpha}c_{k\alpha}^\dagger c_{k\alpha},\nonumber \\
H_B^{(B)}&=\sum_{k,\alpha=L,R}\epsilon_{k\alpha}b_{k\alpha}^\dagger b_{k\alpha},\nonumber \\
\end{align}
where $c_{k\alpha}$ and $c_{k\alpha}^\dagger$ ($b_{k\alpha}$ and $b_{k\alpha}^\dagger$) are the annihilation and creation operators of excitation energy $\epsilon_{k\alpha}$ and quantum number $k$ for fermionic (bosonic) bath $\alpha$. The annihilation and creation operators satisfy usual commutation or anti-commutation relations depending on whether the baths are bosonic or fermionic respectively: $\left[b_{k\alpha},b_{l\eta}^\dagger\right]=\delta_{kl}\delta_{\alpha\eta}$, $\left\{c_{k\alpha},c_{l\eta}^\dagger\right\}=\delta_{kl}\delta_{\alpha\eta}$, where $[\cdots]$ and $\{\cdots\}$ are commutator and anti-commutator respectively. The exchange of particle or energy between the system and the baths is governed by the contact Hamiltonian, $H_C$.  The form of the contact Hamiltonian depends on the specific model considered. 
\section{Adiabatic driving and continuous measurement}
Assuming weak coupling between the measurement probe and the system, the contribution from continuous measurement can be kept at same footing with respect to the baths\cite{jacobs2014}. In fact, it has been observed that continuous measurement of position in harmonic oscillators is equivalent to attaching the system with a bath at zero temperature\cite{jacobs2014}. On the other hand,  the bath can also be viewed as a probe that by continuously monitoring and taking the information away from the system helps the system to relax. Based on above assumptions, the quantum master equation for the reduced density matrix of the system $(\rho)$ is given by
\begin{multline}
\frac{d\rho(t,\tau)}{dt}=-i\left[H_S(\tau),\rho(t,\tau)\right]+\sum_{\alpha=L,R}{\cal L}_{\alpha}(\tau) \rho(t,\tau)\\
+{\cal L}_M(t)\rho(t,\tau),
\label{eq:mas_main}
\end{multline}
where the first term on the right hand side gives the unitary evolution of the density matrix with respect to the system Hamiltonian whereas the second term gives the contribution of the baths to the leading order of system-bath coupling strength (see Ref.~\onlinecite{bibekgreen} for the microscopic derivation of adiabatic quantum master equation). The last term on the right hand side is the quantum measurement contribution.  It can be expressed as\cite{wiseman2009,jacobs2014}
\begin{multline}
{\cal L}_M\rho=\Gamma_M\left[X\rho X^\dagger -\frac{1}{2}\left[X^\dagger X\rho +\rho X^\dagger X\right]\right]\\
+\sqrt{2\Gamma_M}\left[X\rho +\rho X^\dagger-\langle X+X^\dagger\rangle\rho\right]
\frac{dW}{dt},
\label{eq:meas_con}
\end{multline}
where $\Gamma_M$ determines the strength of the measurement and $X$ is the system operator being measured. $dW$ is a stochastic quantity which results from the random nature of measurements. The distribution for $dW$ is Gaussian with zero mean and variance $dt$.  

We have introduced two time scales in Eq.~(\ref{eq:mas_main})(see Ref.~\onlinecite{bibekgreen,pablo} for more details): $t$ is the relaxation time for the system in the presence of baths and the measurement probe.  In other words assuming Markovian dynamics, for $t\rightarrow\infty$ the system goes to the steady state. $\tau$ is associated with the adiabatic driving and runs from 0 to $2\pi/\Omega$ for periodic driving, $\Omega$ is the driving frequency. We consider the driving to be slow enough such that $2\pi \Omega^{-1}\gg t$. This implies $\hbar \Omega$ is the smallest energy scale in the system and $\hbar \Omega \ll \Gamma$, where $\Gamma$ is the average coupling strength between the system and the environment (baths and the measurement probe). 

In the weak system-bath coupling and adiabatic driving limit,  the master equation can be broken down into instantaneous and adiabatic contributions\cite{janine2006,janinedot,bibekgreen,bibekgeo}.  The instantaneous contribution to the master equation is given by
\begin{multline}
\frac{d\rho^{(i)}(t,\tau)}{dt}=-i\left[H_S(\tau),\rho^{(i)}(t,\tau)\right]+\sum_\alpha{\cal L}_\alpha(\tau)\rho^{(i)}(t,\tau)\\
+{\cal L}_M\rho^{(i)}(t,\tau)
\label{eq:mas_ins}
\end{multline}
and the adiabatic correction
\begin{multline}
\frac{d\rho^{(i)}(t,\tau)}{d\vec{U}}\cdot \frac{d\vec{U}}{d\tau}=-i\left[H_S(\tau),\rho^{(a)}(t,\tau)\right]\\
+\sum_\alpha{\cal L}_\alpha(\tau)\rho^{(a)}(t,\tau)+{\cal L}_M\rho^{(a)}(t,\tau),
\label{eq:mas_adia}
\end{multline}
where $\rho=\rho^{(i)}+\rho^{(a)}$, $\rho^{(i)}$ is the instantaneous contribution and $\rho^{(a)}$ is the adiabatic contribution to the density matrix. The expression on the left hand side of Eq.~(\ref{eq:mas_adia}) has two contributions: one associated with the rate of change of instantaneous density matrix as a function of the driving $\tau$ and other due to the change of eigenstates $\vec{U}$ with respect to $\tau$. The later contribution vanishes under the secular approximation\cite{bibekgreen,abiuso2}.The two density matrix satisfy the following normalization condition:
\begin{equation}
{\rm Tr}\left[\rho^{(i)}\right]=1;\;\;{\rm Tr}\left[\rho^{(a)}\right]=0.
\end{equation}
\subsection{Discarding measurement record and the steady state}
\label{sec:no_rec}
In this paper, we will define the steady state only for the case when the measurement record is discarded. Later, we will also study the transient regime where we will take into account the stochasticity associated with quantum measurement process. When the measurement records are discarded, the system reaches the steady state for $t\rightarrow\infty$. The contribution to the system dynamics due to continuous quantum measurement reduces to
\begin{equation}
{\cal L}_M\rho=\Gamma_M \left[X\rho X^\dagger-\frac{1}{2}\left(X^\dagger X\rho+\rho X^\dagger X\right)\right].
\end{equation}
In the steady state, the time derivative of the instantaneous density matrix goes to zero, i.e ${d{\rho}^{(i)}(t\rightarrow \infty,\tau)}/{dt}=0$.  We represent the steady state density matrix as, $\rho^{(i/a)}(\infty,\tau)\equiv\rho^{(i/a)}(\tau)$. The quantum master equations become
\begin{multline}
0=-i\left[H_S(\tau),\rho^{(i)}(\tau)\right]+\sum_\alpha{\cal L}_\alpha(\tau)\rho^{(i)}(\tau)
+{\cal L}_M\rho^{(i)}(\tau)
\label{eq:mas_st_in}
\end{multline}
for the instantaneous contribution and 
\begin{multline}
\frac{d\rho^{(i)}(\tau)}{d\vec{U}}\cdot\frac{d\vec{U}}{d\tau}=-i\left[H_S(\tau),\rho^{(a)}(\tau)\right]\\
+\sum_\alpha{\cal L}_\alpha(\tau)\rho^{(a)}(\tau)+{\cal L}_M\rho^{(a)}(\tau)
\label{eq:mas_st_ad}
\end{multline}
for the adiabatic contribution. 
\section{Heat currents, power and efficiency}
There can be three different contributions to the heat currents: 1) difference of temperature in the left and the right bath, 2) adiabatic driving that pumps heat from one bath to another bath, and 3) quantum measurement which influences the state of the system leading to the change in the heat current flowing into the baths. The heat current flowing out of bath $\alpha$ is given by
\begin{equation}
J_\alpha^{(i/a)}(t,\tau)={\rm Tr}\left[H_S(\tau){\cal L}_\alpha \rho^{(i/a)}(t,\tau)\right],
\label{eq:cur_bath_bm}
\end{equation}
where $J_\alpha^{(i)}$ is the instantaneous heat current and is zero for $T_{\rm L}=T_{\rm R}$ and $\Gamma_{\rm M}=0$ whereas $J_\alpha^{(a)}$ is the adiabatic contribution to the heat current. The effect of quantum measurement is manifested through the density matrix. The heat current flowing out of the measurement probe to the system is given by
\begin{equation}
J_{M}(t,\tau)={\rm Tr}\left[H_S(\tau) {\cal L}_M\rho(t,\tau)\right].
\label{eq:pow_meas}
\end{equation}
The power supplied by the driving is given by
\begin{equation}
P_{D}(t,\tau)=\frac{d\vec{U}}{d\tau}\cdot \vec{\rho}(t,\tau).
\label{eq:pow_dri}
\end{equation}
The setup can be utilized to obtain a variety of thermal machines such as heat engines, refrigerators, heat pumps, thermal accelerators and so on. The choice of thermal machine we obtain depends on the parameter regime considered.  In this paper, we are mostly interested in the process of refrigeration. A refrigerator extracts heat from a cold bath (R) and deposit it into the hot bath (L) using the power provided by the adiabatic driving as well as the quantum measurement. The coefficient of performance for the refrigerator is given by the ratio of the amount of heat extracted from the bath R to the total amount of power supplied to the system, i.e.
\begin{equation}
C(t,\tau)=\frac{J_R(t,\tau)}{P_{D}(t,\tau)+J_{M}(t,\tau)}.
\label{eq:cop}
\end{equation}
where $J_\alpha(t,\tau)=J_\alpha^{(i)}(t,\tau)+J_\alpha^{(a)}(t,\tau)$. The above expression for the coefficient of performance does not include dissipation in the measurement probe. The coefficient of performance in the instantaneous limit is given by
\begin{equation}
C^{(i)}(t,\tau)=\frac{J_R^{(i)}(t,\tau)}{J_{M}^{(i)}(t,\tau)}.
\label{eq:cop_fr}
\end{equation}
Similarly, if the device were to work as a heat engine, it would utilize a part of heat flow due to thermal bias to work against the mechanism of adiabatic driving and quantum measurement. The efficiency of such a device would be given by an inverse ratio\cite{bibekgeo}.

For any observable $O(t,\tau)$ of Eq.~(\ref{eq:cop}), the time averaged counterpart in the steady state is given by
\begin{equation}
\bar{O}=\frac{\Omega}{2\pi}\int_0^{\frac{2\pi}{\Omega}}O(\tau) d\tau.
\label{eq:time_av}
\end{equation}
\subsection{First and the second law of thermodynamics}
Along with the measurement Hamiltonian $(H_{\rm M})$, the total Hamiltonian can be written as
\begin{equation}
H_{\rm tot}(\tau)=H(\tau)+H_{\rm M}.
\end{equation}
The heat currents have a transport as well as a dissipative component. The transport heat current is conservative in nature whereas the dissipative part of the heat current is non-conservative in nature and gives rise to entropy. To study the energy fluxes entering and exiting different parts of the device, we study the time evolution of the Hamiltonian
\begin{equation}
\frac{d\left\langle H_{\rm tot}\right\rangle}{d\tau}=\frac{d\left\langle H\right\rangle}{d\tau}+\frac{d\left\langle H_M\right\rangle}{d\tau},
\label{eq:dynamics}
\end{equation}
where the time evolution would be given by the Heisenberg equation of motion. For instance,
\begin{equation}
\frac{d\left\langle H\right\rangle}{d\tau}=i\left\langle\left[H_{\rm tot},H\right] \right\rangle +\left\langle\frac{\partial H_S}{\partial \tau} \right\rangle.
\end{equation}
In the steady state, the dynamics for the bath and contact Hamiltonian would be given by
\begin{align}
\frac{d\left\langle H_B\right\rangle}{d\tau}=i\left\langle\left[H_{\rm tot},H_B\right] \right\rangle& =\sum_\alpha J_{\alpha}(\tau),\nonumber \\
\frac{d\left\langle H_C\right\rangle}{d\tau}=i\left\langle\left[H_{\rm tot},H_C\right] \right\rangle& =\sum_\alpha J_{C,\alpha}(\tau),
\end{align}
where the instantaneous and adiabatic components of $J_{\alpha}(\tau)$ will be given by steady state solutions of Eq.~(\ref{eq:cur_bath_bm}) in the Born-Markov approximation. Similarly we can identify the power provided by the driving as well as the measurement probe as
\begin{align}
P_D(\tau)&=\left\langle \frac{\partial H_S}{\partial \tau}\right\rangle,\nonumber \\
J_M(\tau)&=i\left\langle \left[H_S,H_M\right]\right\rangle.\nonumber \\
\end{align}
Note that when the system and the measurement Hamiltonians commute, the power provided by the measurement probe to the system vanishes.  Hence, only the measurement of observables which do not commute with the system Hamiltonian can power measurement based thermal devices. 

Similar to the case of heat currents, the expression for the powers in the Born-Markov approximation would be given by the steady state solutions of Eqs.~(\ref{eq:pow_dri}) and (\ref{eq:pow_meas}). Substituting above relations for powers and heat currents in Eq.~(\ref{eq:dynamics}), we obtain
\begin{equation}
0=J_S(\tau)+\sum_\alpha\left(J_\alpha(\tau)+J_{c,\alpha}(\tau)\right)+J_M(\tau)
\label{eq:first_law}
\end{equation}
which gives the first law of thermodynamics.  We defined, $J_{S}(\tau)=i\left\langle\left[H_{\rm tot},H_S\right]\right\rangle$ and used $\frac{d\left\langle H_{\rm tot}\right\rangle}{d\tau}=\left\langle \frac{\partial H_S}{\partial \tau}\right\rangle=P_D$. The heat current stored in the contact region goes to zero on time-averaging.  Eq.~(\ref{eq:first_law}) reduces to
\begin{equation}
0=\bar{J}_S+\sum_\alpha\bar{J}_\alpha+\bar{J}_M
\end{equation}
The power dissipated by the driving and the measurement probe are non-conservative in nature and in general gives rise to entropy production. Therefore, when both baths as well as the measurement probe are kept at temperature $T$, the time-average entropy production rate would be given by
\begin{equation}
T\dot{\bar{S}}=\bar{P}_D+\bar{P}_M,
\end{equation}
where $\bar{P}_M$ is the power dissipated due to measurement which is neglected while calculating the coefficient of performance in Eqs.~(\ref{eq:cop}) and (\ref{eq:cop_fr}). Since, both $\bar{P}_D$ and $\bar{P}_M$ are greater than or equal to zero, we obtain $\dot{\bar{S}}\geq 0$ giving the second law of thermodynamics. If the measurement probe and the two baths have different temperatures, which is the case in general,  the total entropy production will be given in terms of heat currents, the corresponding temperatures and the dissipated power due to driving and measurement\cite{strasberg,bibekgeo}. The formulation used in this paper only addresses the change in the dynamics of the system due to the measurement probe but not vice-versa and hence is inadequate to study the dissipation in the measurement probe.  
\subsection{Geometric nature of the work done by quantum measurement}
\label{sec:geo_meas}
\begin{figure}[t]
\includegraphics[width=\columnwidth]{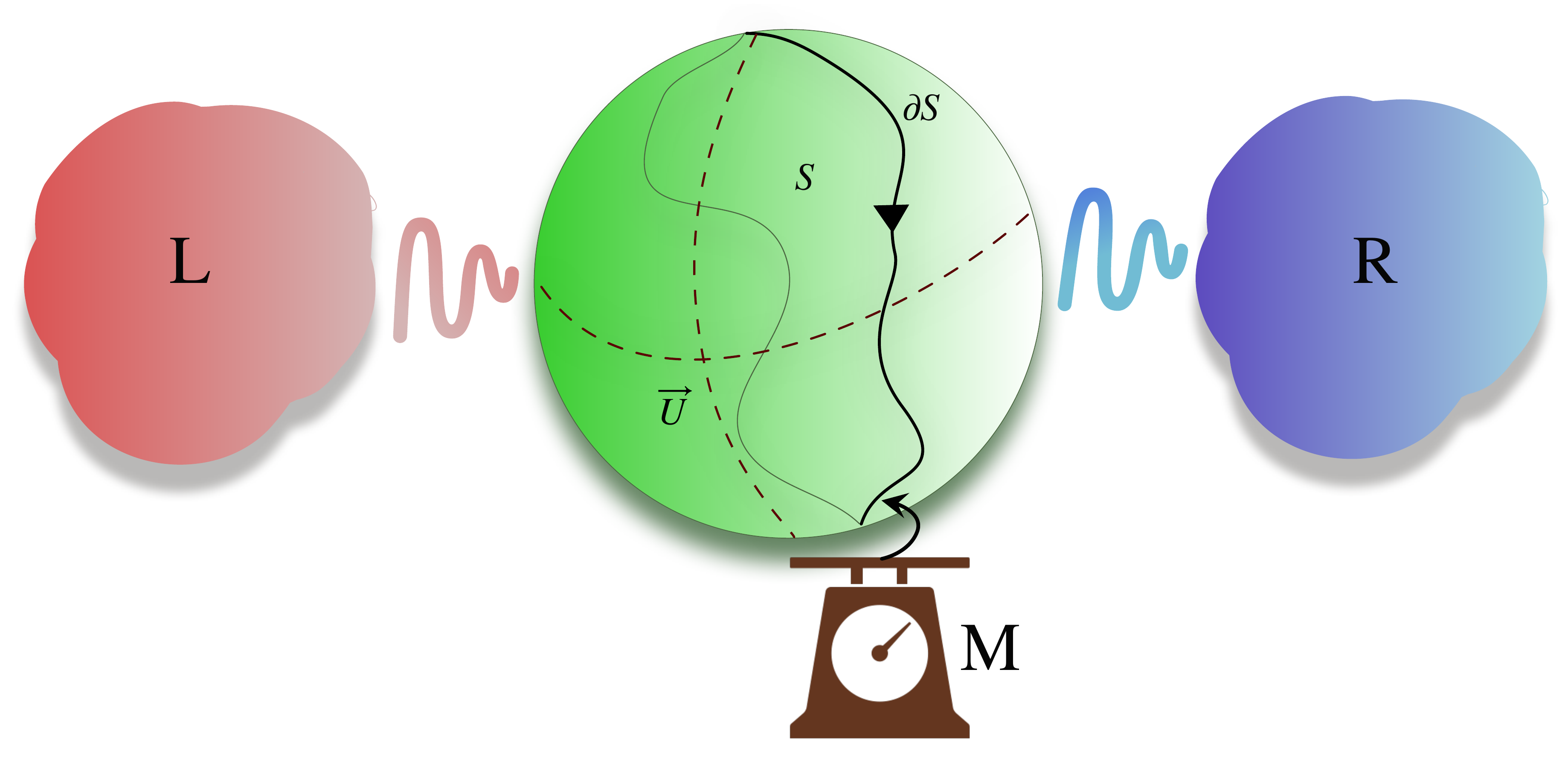}	
	\caption{The contour ($\partial S$) traversed by $\vec{\Lambda}$ in the parameter space defined by $\vec{U}$. The adiabatic work done by the continuous measurement is given by the flux of curl of $\vec{\Lambda}$ across surface $S$.}
	\label{fig:setupgeo}
\end{figure}
When a quantum system is driven adiabatically in a cycle, the system's final state acquires a gauge-invariant phase which is geometric in nature\cite{berry1984}. Based on aforementioned arguments, it was recently shown that different thermodynamic quantities associated with adiabatically driven open quantum systems can have geometric contribution\cite{bibekgeo,pablo}. Different thermodynamic quantities associated with adiabatically driven quantum systems in the linear response regime can be described in terms of thermal geometric tensor\cite{bibekgeo,pablo}.  Heat-to-work conversion which lies at the heart of the performance of thermal machines is given by the anti-symmetric component of the thermal geometric tensor. Furthermore, the anti-symmetric component have the structure of Berry curvature and depends only on the geometry of the trajectory traversed in the parameter space. On the other hand, the symmetric component of the thermal geometric tensor do not have a structure of Berry curvature but has a geometric interpretation in terms of thermodynamic length.  The symmetric components are dissipative in nature and gives rise to entropy production.  Consequently, optimization protocols in adiabatically driven system is determined by the best trade-off between the anit-symmetric and symmetric components of the thermal geometric tensor\cite{pablo}. From now on, we will use the term `geometric' only for the terms having the structure of Berry curvature.

It has been experimentally observed that a sequence of weak measurements continuously varying in measurement strength can induce a geometric phase akin to the one obtained with adiabatic driving \cite{cho2019}. Futhermore, topological transitions in measurement-induced geometric phases when one gradually varies the measurement strength was recently studied theoretically in Ref.~\onlinecite{gebhart2020}.
The aforementioned observations motivates the study of geometric characteristics of measurement based thermal devices. 

Neglecting the off-diagonal terms of the density matrix, the commutator in Eqs.~(\ref{eq:mas_st_in}) and (\ref{eq:mas_st_ad}) vanishes. The adiabatic contribution to the density matrix can be written in the following form
\begin{equation}
\rho^{(a)}(\tau)=\left(\sum_\alpha{\cal L}_\alpha(\vec{U})+{\cal L}_M\right)^{-1}\frac{d\rho^{(i)}(\tau)}{d\vec{U}}\cdot\frac{d\vec{U}}{d\tau}.
\label{eq:prob_ad}
\end{equation}
Note that the off-diagonal terms were neglected only for the sake of simplicity. The results in this section hold even when quantum coherence is taken into account. Plugging Eq.~(\ref{eq:prob_ad}) in Eq.~(\ref{eq:pow_meas}), the adiabatic contribution to the heat current flowing out of measurement probe is given by
\begin{equation}
J_M^{(a)}(\tau)=\\
{\rm Tr}\left[H_S(\vec{U}){\cal L}_M \left[\vec{R}(\vec{U})\rho^{(i)}(\tau)\right]\cdot \frac{d\vec{U}}{d\tau}\right],
\label{eq:pow_meas_ad}
\end{equation}
where 
\begin{equation}
\vec{R}(\vec{U})=\left(\sum_\alpha{\cal L}_\alpha(\vec{U})+{\cal L}_M\right)^{-1}\frac{d}{d\vec{U}}.
\end{equation}
Defining $\vec{\Lambda}={\rm Tr}\left[H_S(\vec{U}){\cal L}_M \left[\vec{R}(\vec{U})\rho^{(i)}(\tau)\right]\right]$, the time-average of Eq.~(\ref{eq:pow_meas_ad}) expressed as
\begin{equation}
\frac{2\pi}{\Omega}\bar{J}_M^{(a)}=
\int_{\partial S}\vec{\Lambda}
\cdot {d\vec{U}}
\end{equation}
is a geometric quantity which implies that the amount of work done by continuous measurement per driving period do not depend on the details of the driving but only on the trajectory traversed ($\partial S$) by $\vec{\Lambda}$ in the parameter space defined by $\vec{U}$ (see Fig.~\ref{fig:setupgeo}).

We define the adiabatic work done by measurement through $2\pi \bar{J}_{M}^{(a)}=\Omega \bar{W}_M^{(a)}$. Using the Stoke's theorem, one can express the adiabatic work performed as a surface integral given by
\begin{equation}
\bar{W}_M^{(a)}=\int_{\partial S} \vec{\Lambda}.d\vec{U}=\int_S\left(\nabla_U\times \vec{\Lambda}\right)\cdot d\vec{S},
\end{equation}
where $S$ is the surface defined in the $\vec{U}$ space whose bounday is given by the control trajectory ($\partial S$).  In the geometrical interpretation, the adiabatic work done by the continuous quantum measurement process is given by the flux of the curl of $\vec{\Lambda}$ across surface $S$, provided $\partial S$ is a simple closed curve and $\vec{\Lambda}$ has continuous first order partial derivatives in the region defined by $S$.  For $\Delta T=0$ and in the absence of measurement, the work done by the driving force has no geometric component associated with it and is entirely dissipative in nature\cite{bibekgeo}. Instead, the work done by the continuous measurement has both geometric ($\bar{W}_M^{(a)}$) as well as dissipative components ($\bar{W}_M^{(i)}$) even for $\Delta T=0$. The geometric optimization protocol and heat-to-work conversion is rooted in the geometric components of work done to the system. As observed in Ref.~\onlinecite{pablo}, the optimization is based on finding a protocol with a maximum ratio between the geometric and dissipative components of the work done to the system.  For the case of $\Delta T=0$, this ratio ($\kappa$) in our case would be given by
\begin{equation}
\kappa=\frac{\bar{J}_M^{(a)}}{\bar{J}_M^{(i)}+\bar{P}_D}.
\end{equation}

\section{Coupled Qubits and Coupled Quantum dots}
In order to illustrate the ideas developed in the previous sections,  we consider two different examples: a coupled qubit system in contact with bosonic reservoirs and a coupled quantum dot system in contact with fermionic reservoirs. One of the qubits (quantum dots) is coupled to the bath L and the other one to the bath R. The Hamiltonian for coupled qubit can be written as
\begin{equation}
H_{S,Qbt}=\epsilon_L\hat{\pi}_{LL}+\epsilon_R \hat{\pi}_{RR}+{g}\left(\hat{\pi}_{LR}+\hat{\pi}_{RL}\right),
\end{equation}
where $\pi_{jk}=|j\rangle \langle k|$ is the outer product operator onto the excited state of qubit $j(k)$.  We choose a common ground state $0$ for both qubits, and fix it's energy to zero, $\epsilon_0=0$.  
The Hamiltonian for the coupled quantum dot system is given by
\begin{equation}
H_{S,QD}=E_L a_L^\dagger a_L +E_R a_R^\dagger a_R +{G}\left(a_L^\dagger a_R + a_{R}^\dagger a_L\right),
\end{equation}
where $a_j(a_j^\dagger)$ are the creation annihilation operators for quantum dot $j$ and $E_j$ is the bare energy of the quantum dot $j$. The fermionic annihilation and creation operators for the quantum dots (allowing no more than one electron in the quantum dots) can be written in terms of outer product operators as
\begin{equation}
a_j = |0\rangle\langle j| ;\;\;\; a_j^\dagger = |j\rangle\langle 0|
\end{equation}
such that the system Hamiltonian becomes
\begin{equation}
 H_{S,QD}=E_L\hat{\pi}_{LL}+E_R \hat{\pi}_{RR}+{G}\left(\hat{\pi}_{LR}+\hat{\pi}_{RL}\right),
\end{equation}
We observe that in terms of outer product operators $\pi_{mn}$, the system Hamiltonian for the coupled quantum dot and coupled qubit setup are equivalent. Hence we will denote by $H_S$ both the system Hamiltonians. We will consider fermionic baths for coupled quantum dots and bosonic baths for the case of qubits. The Hamiltonian for the baths are given in Sec.~(\ref{sec:model}).
We will consider two different coupling Hamiltonians for the bosonic case. The linear coupling Hamiltonian,
\begin{equation}
H_C=\sum_{k,\alpha=L,R}V_{k\alpha}\left[\pi_{\alpha 0}b_{k\alpha}+\pi_{0\alpha}b_{k\alpha}^\dagger\right].
\label{eq:coup_bos_lin}
\end{equation}
In some cases, we will also consider non-linear coupling between the system and bath R given by\cite{bibekrect}
\begin{equation}
H_{C,R}=\sum_{k}V_{kR}\left[\pi_{R 0}\,b_{kR}^2+\pi_{0R}\left(b_{kR}^{\dagger}\right)^2\right].
\label{eq:coup_bos_nonlin}
\end{equation}
For the fermionic case, the contact Hamiltonian is
\begin{equation}
H_C=\sum_{k,\alpha=L,R}V_{k\alpha}a_{\alpha}^\dagger c_{k\alpha}+h.c.
\label{eq:coup_fer}
\end{equation}
We can diagonalize the system Hamiltonian through a suitable change of basis $|L\rangle = \cos \theta |+\rangle -\sin\theta |-\rangle$ and $|R\rangle = \sin\theta |+\rangle +\cos\theta |-\rangle$
where for the case of coupled qubits $\theta = \frac{1}{2}\tan^{-1}\left(\frac{2g}{\epsilon_L-\epsilon_R}\right)$,
$
\sin 2\theta =2g/h;\; \cos2\theta = \Delta/h,$ 
where $h=\sqrt{4g^2+\Delta^2}$ and $\Delta=\epsilon_L-\epsilon_R$. Similar relations can be obtained for the coupled quantum dot case simply by replacing $\epsilon_j$ with $E_j$ and $g$ with $G$ as the two system Hamiltonians are equivalent. In addition,we obtain
\begin{equation}
\sin\theta= \sqrt{\frac{h-\Delta}{2h}};\;\cos\theta= \sqrt{\frac{h+\Delta}{2h}}.
\end{equation} In the new basis the system Hamiltonian reads
\begin{equation}
H_S=\epsilon_+ |+\rangle \langle +| +\epsilon_- |-\rangle \langle -|,
\label{eq:hamil_diag}
\end{equation}
where $
\epsilon_{\pm}=({\epsilon_L
+\epsilon_R})/{2}\pm {h}/{2}$ for the case of qubits and $
\epsilon_{\pm}=({E_L
+E_R})/{2}\pm {h}/{2}$ for the case of quantum dots.  In Eq.~(\ref{eq:hamil_diag}), we do not include the term representing the state ($|0\rangle$) when both quantum dots are empty or both qubits are in the ground state assuming the energy associated with state $|0\rangle$ is zero. The contact Hamiltonian gets modified accordingly. In the following subsections, we will study the dynamics considering only the diagonal terms of the density matrix,  the calculations with the off-diagonal terms is presented in App.~\ref{app:mas}. 
\subsection{Dynamics induced by bath and continuous measurement}
\label{sec:dyn_diag}
We will use the global quantum master equations to study the transfer of heat and dissipation in the aforementioned two setups. We will investigate only the regime where global master equations are valid, i.e. $\Gamma_\alpha \lesssim g,\;|\epsilon_{\rm L}-\epsilon_{\rm R}|\gg g$ for the case of quantum dots and $\Gamma_\alpha (E_\alpha)\lesssim G,\;|E_{\rm L}-E_{\rm R}|\gg G$ for the case of qubits\cite{sabrina}. If we disregard the off-diagonal terms of the density matrix,  the density matrix can be written in a simple vector form $\rho = \begin{bmatrix}\rho_{00}&\rho_{++}&\rho_{--}\end{bmatrix}^T$, where $\rho_{jj}=\langle j|\rho|j\rangle$ is the population of the eigen-state $j$ and $\rho_{jj}=\rho_{jj}^{(f)}+\rho_{jj}^{(a)}$. Moreover, the Lindbladian associated with the bath $\alpha$ is given by
\begin{equation}
{\cal L}_\alpha =\begin{bmatrix}
-\left(\gamma_{\alpha,0+}+\gamma_{\alpha,0-}\right) & \gamma_{\alpha,+0} &\gamma_{\alpha,-0}\\
 \gamma_{\alpha,0+} & -\gamma_{\alpha,+0} & 0\\
\gamma_{\alpha,0-} & 0 &-\gamma_{\alpha,-0},
\end{bmatrix}
\label{eq:master_eq}
\end{equation}
where $\gamma_{\alpha,0m}$ and $\gamma_{\alpha,m0}$ are the transition rates for going from state 0 to state $m$ and from state $m$ to state $0$ respectively. For linear system-bath coupling, they are defined as
\begin{align}
&\gamma_{\alpha,m0}=\hbar^{-1}\lambda_{\alpha,0m}\Gamma_\alpha(\epsilon_{m0})\Big(1\pm n_\alpha(\epsilon_{m0})\Big),\nonumber \\
&\gamma_{\alpha,0m}=\hbar^{-1}\lambda_{\alpha,0m}\Gamma_\alpha(\epsilon_{m0})n_\alpha(\epsilon_{m0}),
\end{align}
where $\lambda_{L,0+}=\lambda_{R,0-}=\cos^2\theta$, $\lambda_{R,0+}=\lambda_{L,0-}=\sin^2\theta$, $\epsilon_{m0}=\epsilon_m-\epsilon_0$, $n_\alpha(\omega)$ is the Fermi-Dirac or Bose-Einstein distribution depending on whether we are considering coupled quantum dot or coupled qubit system respectively and $\Gamma_{\alpha}(\omega)$ is the spectral density for bath $\alpha$ defined as
\begin{equation}
\Gamma_\alpha(\omega)=2\pi \sum_k |V_{k\alpha}|^2\delta(\omega-\epsilon_{k\alpha}).
\end{equation}
For fermionic baths, we will consider characterless baths, i.e. $\Gamma_{\alpha}(\omega)=\Gamma_\alpha(0)\equiv \Gamma_\alpha$ whereas for the Bosonic baths we will consider Ohmic baths, $\Gamma_{\alpha}(\omega)=\Upsilon_\alpha \omega e^{-\omega/\omega_C}$, where $\omega_C$ is the cut-off frequency.

The measurement operator is chosen to be $X=|R\rangle\langle R|$. Taking only the diagonal terms of the density matrix, we have
\begin{equation}
\langle X\rangle = \sin^2\theta \rho_{++}+\cos^2\theta \rho_{--}.
\end{equation}
Moreover,
\begin{align}
\langle 0 | {\cal L}_M\rho|0\rangle & = -2\sqrt{2\Gamma_M}\langle X\rangle \rho_{00} \frac{\Delta W}{\Delta t},\nonumber \\
\langle + | {\cal L}_M\rho|+\rangle & =\sin^2\theta \cos^2\theta \left(\rho_{--}-\rho_{++}\right)\nonumber \\
&\;\;\;\;\;\;\;\;+2\sqrt{2\Gamma_M}\left(\sin^2\theta \rho_{++}-\langle X\rangle \rho_{++}\right) \frac{\Delta W}{\Delta t}.
\end{align}
We will consider a Gaussian distribution for $W$ with zero mean and variance $\Delta t$. For each small time step $\Delta t$, $\Delta W$ would be selected randomly from the Gaussian distribution. The sequence of values of $dW$ gives the noise realisation for the particular trajectory the system follows.
\subsection{Heat current and coefficient of performance}
The time resolved density matrix calculated using Eqs.~(\ref{eq:mas_ins}) and (\ref{eq:mas_adia}) along with Lindbladian calculated in the previous section fully describes the dynamics of the system. For linear system-bath coupling, the heat current flowing out of bath $\alpha$ can be expressed in terms of density matrix and transition rates as
\begin{multline}
J_\alpha^{(i/a)}(t,\tau)=\sum_m \epsilon_{m0}(\tau)\Big[-\gamma_{\alpha,m0}(\tau)\rho_{mm}^{(i/a)}(t,\tau)\\
+\gamma_{\alpha,0m}(\tau)\rho_{00}^{(i/a)}(t,\tau)\Big].
\label{eq:curr_diag}
\end{multline}
%The time averaged heat current in the steady state 
%\begin{equation}
%\bar{J}_\alpha^{(i/a)}=\frac{2\pi}{\Omega}\int_0^{2\pi/\Omega} {J}_\alpha^{(i/a)}(\infty,\tau)d\tau.
%\end{equation}
In the steady state, the power provided by the continous measurement is given by
\begin{equation}
J_M(\tau)=
\sin^2\theta\cos^2\theta\,\Gamma_M\left(\epsilon_+(\tau)-\epsilon_-(\tau)\right)\left(\rho_{--}-\rho_{++}\right).
\end{equation}
Similarly the power provided by the driving source is given by
\begin{equation}
P_D(\tau)=\sum_m \frac{d\epsilon_m}{d\tau}\rho_{mm}(\tau).
\end{equation}
The time averaged heat currents and powers can be calculated using Eq.~(\ref{eq:time_av}). Using Eqs.~(\ref{eq:cop}) and (\ref{eq:cop_fr}), one can calculate the total and the instantaneous component of the coefficient of performance respectively.
\section{Results: Coupled quantum dots}
In this section, we will present some numerical results for the coupled quantum dot setup mentioned above.  We will study two different conditions: 1) transient dynamics for the undriven case and 2) effect of continuous measurement on steady state adiabatic quantum thermal machines.
\subsection{Time-independent case}
\begin{figure}[!htb]
\includegraphics[width=\columnwidth]{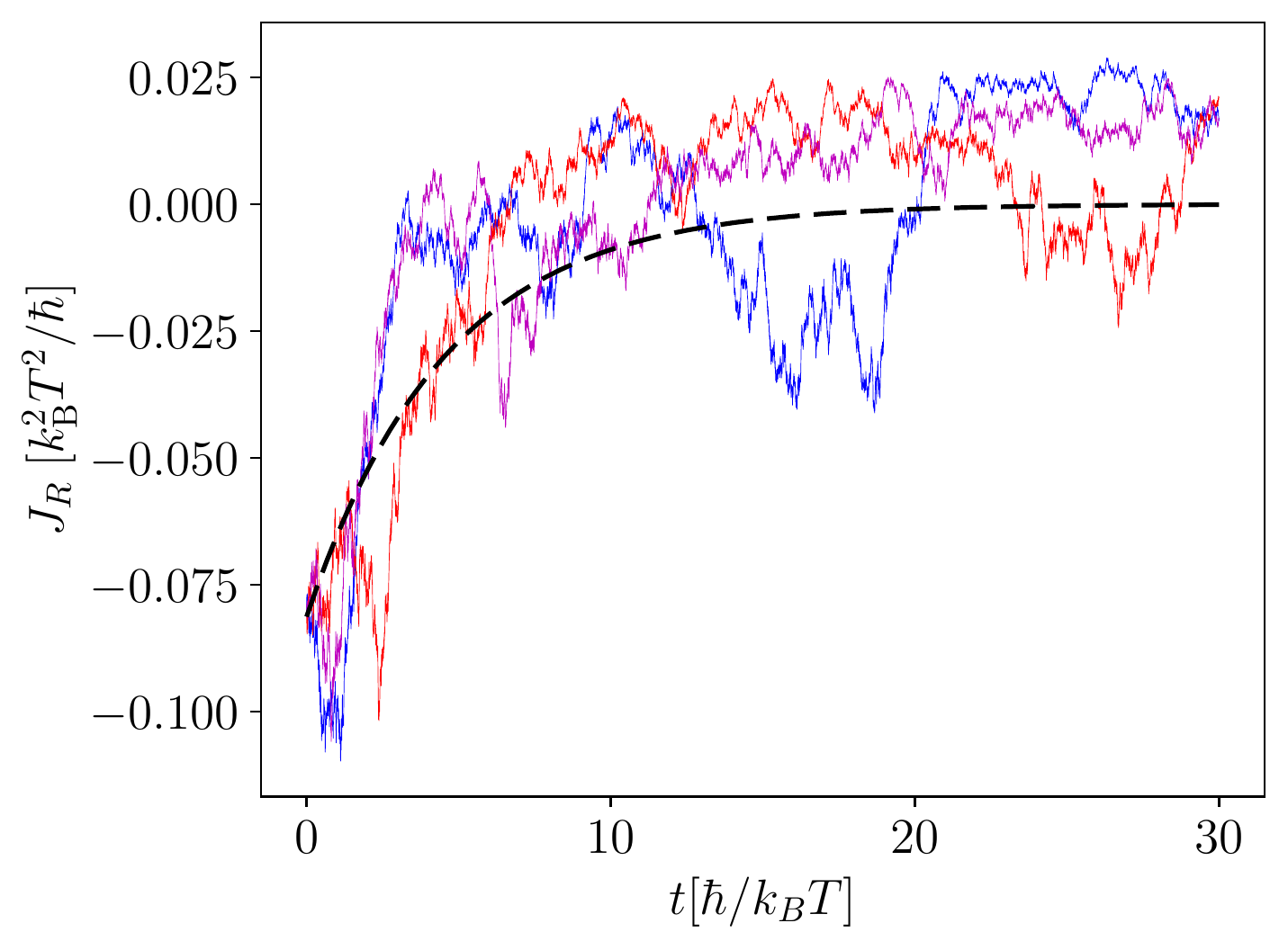}	
	\caption{Heat current as a function of time for time-independent case when the measurement records are discarded (dashed black curve) and when the measurement records are taken or the stochastic term in Eq.~(\ref{eq:meas_con}) is taken into account (solid magenta, blue and red curves). Parameters: $\Gamma_L=\Gamma_R=0.2 k_{\rm B}T$, $k_{\rm B}T_L=k_{\rm B}T_R=k_{\rm B}T=1$, $E_L=2 k_{\rm B}T$, $E_R=3k_{\rm B}T$, $G=0.1k_{\rm B}T$, $\Gamma_M=0.01k_{\rm B}T$, $\Delta t=0.005\hbar/k_{\rm B}T$. The initial state was taken as $\rho_{00}(0)=0.5$, $\rho_{++}(0)=0.2$, $\rho_{--}(0)=0.3$.}
	\label{fig:transient}
\end{figure}
We study the transient dynamics for static coupled quantum dots system both discarding the measurement record as well as keeping the measurement record. For similar set of parameters, we plot the heat current extracted from the right bath as a function of time ($t$) for the aforementioned two cases using Eq.~(\ref{eq:curr_diag}) . In Fig.~\ref{fig:transient}, the black dashed curve is obtained by averaging over the measurement record and the solid curves are obtained using the stochastic master equation. We observe that system reaches the steady state on average for $t\rightarrow \infty$.  For the particular choice of parameters, we observe that heating dominates over the cooling effect on average (black dashed curve). However, if we consider only one realisation of the measurement record, cooling can be obtained for a period of time (see magenta and blue curves for $20<t\leq 30$). One important aspect of keeping the measurement record is that one can resort to quantum feedback loops to suitably tune the dynamics and obtain more efficient thermal machines. For instance, depending on the outcome of measurement one can apply a suitable unitary quantum gate to the system achieving enhanced cooling effects\cite{campisi}. 
\begin{figure}[!htb]
\includegraphics[width=\columnwidth]{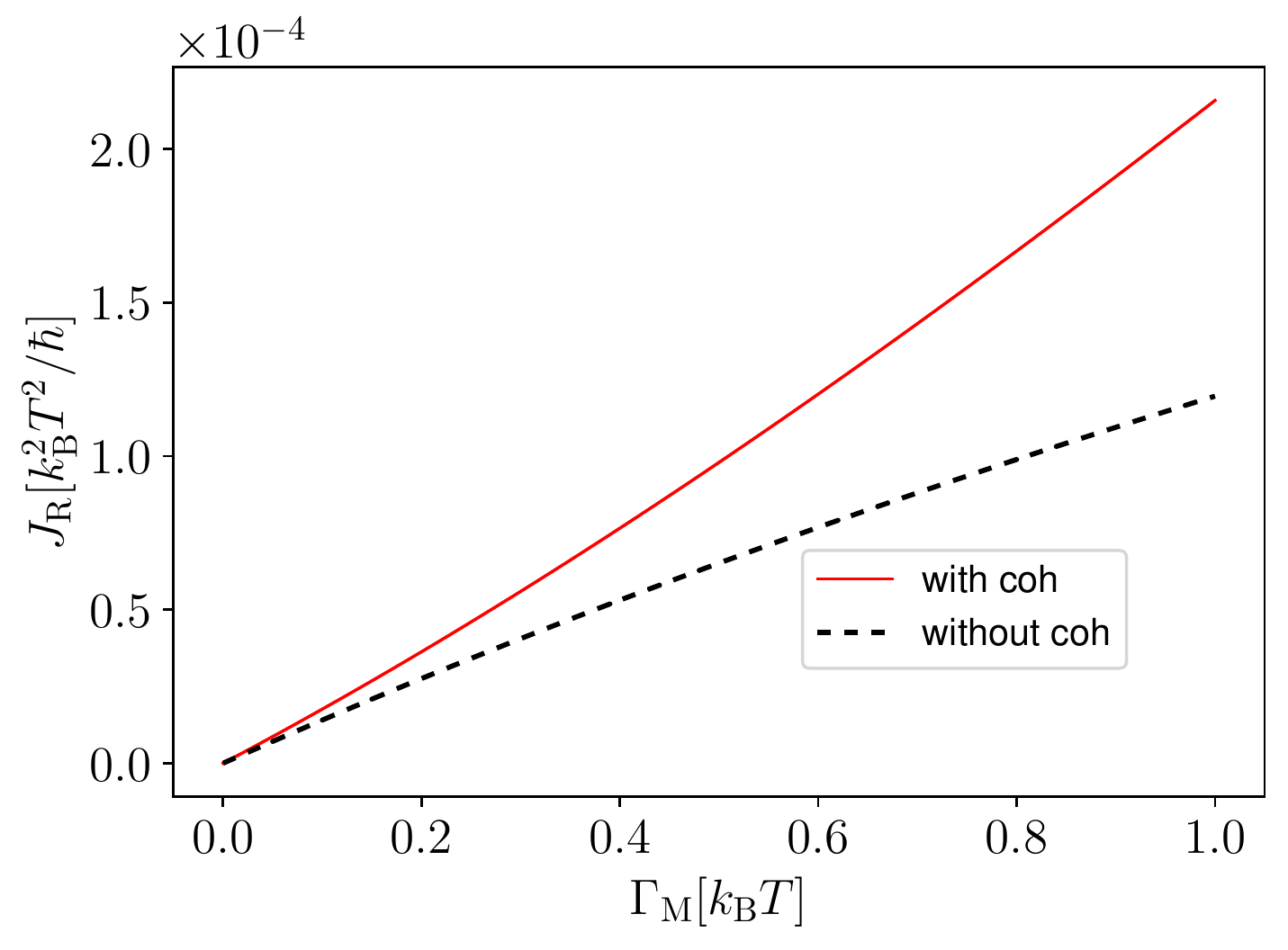}	
	\caption{Steady state cooling as a function of measurement strength with (solid red curve) and without (dashed black curve) eigen-state coherence in coupled quantum dot system. Parameters: $\Gamma_L=\Gamma_R=0.2 k_{\rm B}T$, $G=0.5 k_{\rm B}T$, $E_L=4 k_{\rm B}T$, $E_R=0.15 k_{\rm B}T$}
	\label{fig:coh}
\end{figure}

In Fig.~\ref{fig:coh}, we study the effect of quantum coherence on measurement powered refrigeration. We plot the extracted heat current as a function of measurement strength. The solid red line is obtained when the master equation takes into account both the diagonal as well as off-diagonal terms of the density matrix (see App.~\ref{app:mas}) whereas the dashed black line is obtained using only the diagonal terms of the density matrix as illustrated in Sec.~\ref{sec:dyn_diag}. We observe that for larger values of measurement strength, quantum coherence can significantly enhance the cooling effect. This is in contrast to what is observed in adiabatic quantum refrigerators where quantum coherence usually leads to decrease in cooling\cite{brandner2020}. Quantum coherence can improve cooling of adiabatic quantum refrigerators only in highly non-equilibrium situations\cite{uzdin2015,brandner2017}. However, in our current setup we observe that quantum coherence can improve cooling effect even when $T_{\rm L}=T_{\rm R}$.

\subsection{Driven case}
The quantum dots are driven adiabatically following the protocol: $E_L=E_{0,L}+E_{1,L}\cos\left(\Omega t\right)$ and $E_R=E_{0,R}+E_{1,R}\cos\left(\Omega t+\phi\right)$. The phase difference of $\phi$ is required to pump heat from one bath to another bath.  The amount of heat pumped per cycle depends on the contour traversed on the parameter space but not on the speed of driving. In the presence of continuous measurement, there are two contributions to the heat current 1) instantaneous component which goes to zero in the absence of measurement for $T_{\rm L}=T_{\rm R}$ and 2) adiabatic component which is geometric in nature. The total heat current flowing out of a bath is given by the sum of aforementioned two components.
\begin{figure}[!htb]
\includegraphics[width=0.9\columnwidth]{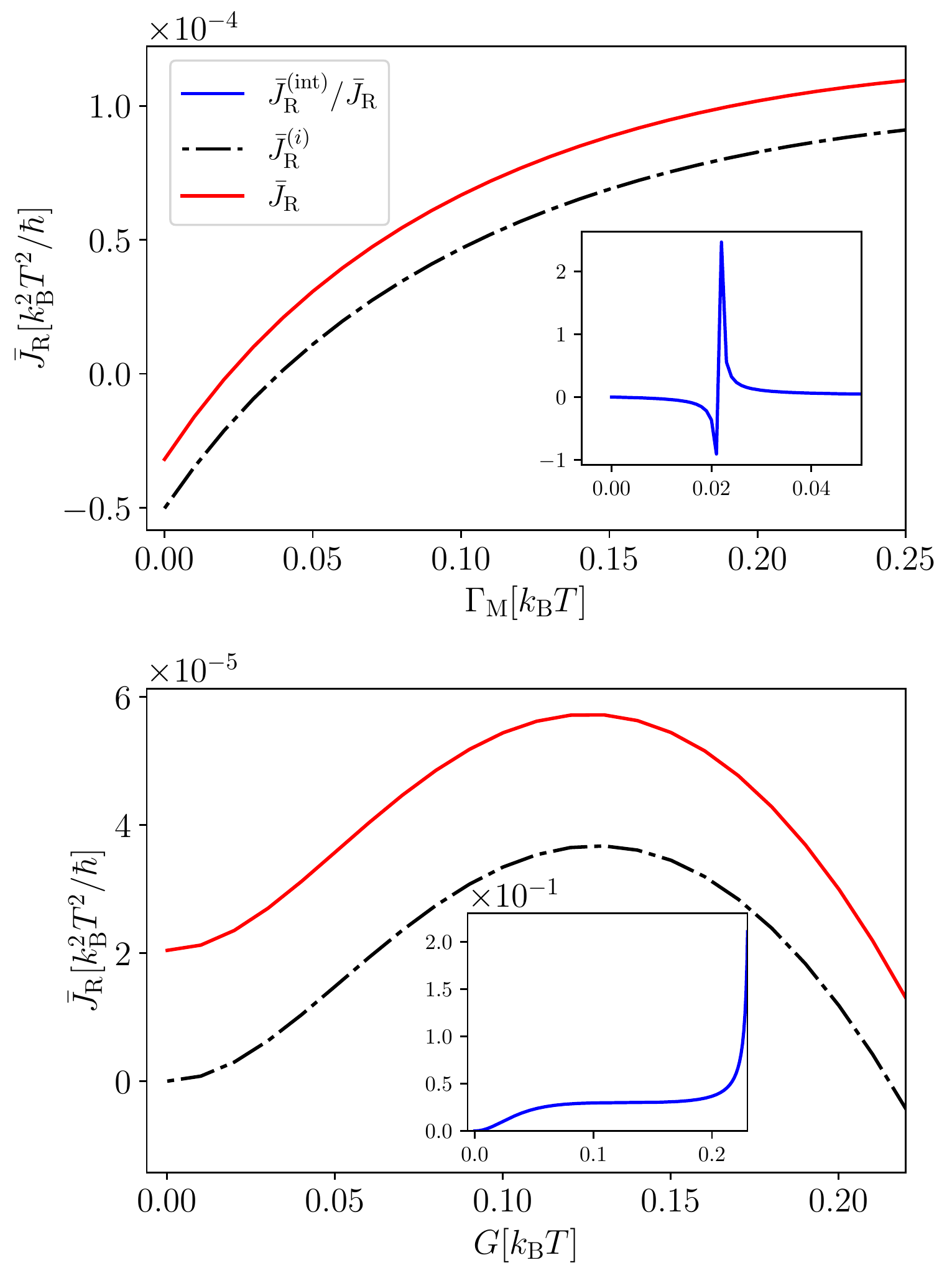}	
	\caption{Cooling as a function of $\Gamma_M$ (top panel) for $G=0.15 k_{\rm B}T$ and inter-dot coupling strength (bottom panel) for $\Gamma_M=0.08 k_{\rm B}T$ in adiabatically driven quantum dots. Parameters: $\Gamma_L=\Gamma_R=0.05 k_{\rm B}T$, $E_L=1.5k_{\rm B}T+0.2k_{\rm B}T\cos\left(0.005k_{\rm B}T t/\hbar\right)$, $E_R=0.3k_{\rm B}T+k_{\rm B}T\cos\left(0.005k_{\rm B}T t/\hbar+\pi/2\right)$, $(T_L+T_R)/2=T$ and $\Delta T=T_L-T_R=0.05 T$.}
	\label{fig:Jiia}
\end{figure}

In the top panel of Fig.~\ref{fig:Jiia}, we study the heat current as a function of measurement strength.  The black dashed curve is for the instantaneous case and the red curve is the total heat current extracted from the right reservoir. We observe that the adiabatic heat current shows a maximum as a function of $\Gamma_M$ (for $\Gamma_{\rm M}\approx 0.1 k_{\rm B}T$).  Hence, a boost in adiabatic heat current is obtained via continuous monitoring of system operator $X$. Although the magnitude of adiabatic heat current is smaller compared to the instantaneous heat current, the adiabatic contribution is always positive in the entire range of $\Gamma_M$ considered and provides a boost to the measurement induced instantaneous heat current (see the black dashed and solid red curves).  Furthermore, we can define the interplay between the two power sources through $\bar{J}_{\rm R}^{\rm (int)}=\bar{J}_{\rm R}-\bar{J}_{\rm R}^{\rm (i)}-\bar{J}_{\rm R}^{\rm (a)}[\Gamma_{\rm M}=0]=\bar{J}_{\rm R}^{\rm (a)}-\bar{J}_{\rm R}^{\rm (a)}[\Gamma_{\rm M}=0]$. Since both $\bar{J}_{\rm R}^{\rm (a)}$ and $\bar{J}_{\rm R}^{\rm (a)}[\Gamma_{\rm M}=0]$ are geometric quantities, the interplay term $\bar{J}_{\rm R}^{\rm (int)}$ is geometric in nature as well. We plot $\bar{J}_{\rm R}^{\rm (int)}$ as a function of $\Gamma_{\rm M}$ in the inset of top panel of Fig.~\ref{fig:Jiia}. We observe that $\bar{J}_{\rm R}^{\rm (int)}$ is positive in the cooling regime ($\Gamma_{\rm M}\geq 0.022 k_{\rm B}T$) and has a maximum in the regime where $\bar{J}_{\rm R}^{\rm (i)}\approx -\bar{J}_{\rm R}^{\rm (a)}$. Hence, we can state that the extracted power due to the combined effect is larger than the sum of individual contributions ($\bar{J}_{\rm R}^{\rm (a)}|_{\Gamma_M=0}+\bar{J}_{\rm R}^{\rm (i)}|_{\Gamma_M\neq 0}$) in the parameter regime considered. Similar observation can be made from the bottom panel of Fig.~{\ref{fig:Jiia}} where we plot the extracted heat current as a function of inter-dot coupling. For $G=0$, the instantaneous heat current vanishes and total heat current becomes equal to the adiabatically pumped heat current. In addition, we observe that the total extracted heat current shows a peak before reducing into the heating regime. For very large values of inter-dot coupling, even the contribution from adiabatic driving goes to the heating regime (not shown in the figure). However, the direction of heat current depends strongly on the driving protocol considered.  Interestingly, there is a finite and positive adiabatic heat current flowing even when the inter-dot coupling $(G)$ goes to zero. We use the global master equation to study the dynamics which is not suited for the regime, $G\ll \Gamma_\alpha,\hbar \Omega$ giving inaccurate results. Since for $G=0$, the system reduces to a single qubit in contact with bath $C$ and one cannot extract heat from a single bath just by driving the system attached to it. We observe a $\approx 2\%$ enhancement due to interplay between adiabatic driving and continuous measurement in the regime where the cooling effect is maximum (see the blue curve in the inset). The interplay term has a stronger effect for $G>0.2 k_{\rm B}T$ where cooling is small.

\begin{figure}[!htb]
\includegraphics[width=0.9\columnwidth]{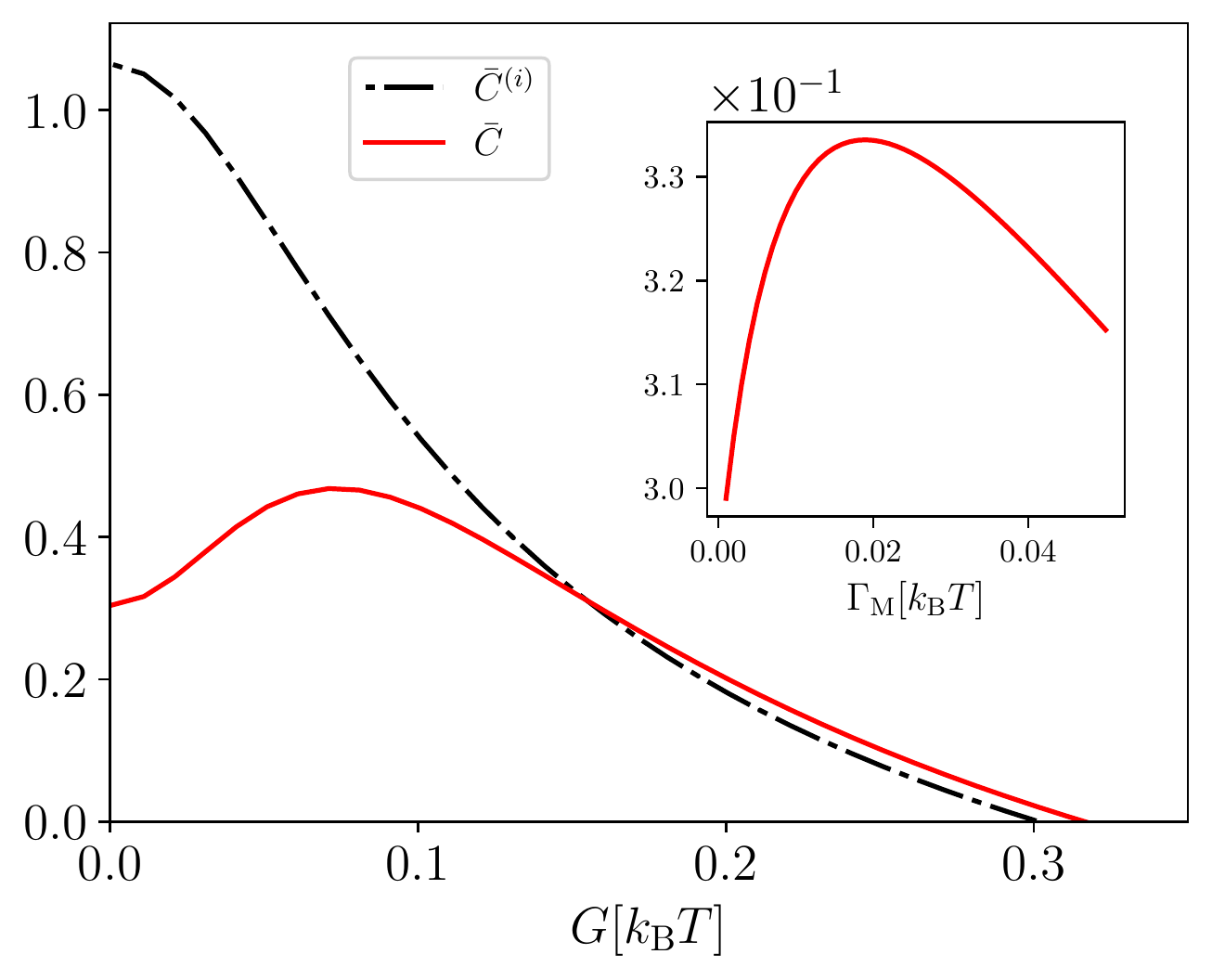}	
	\caption{Instantaneous (black dashed curve) and total (red curve) coefficient of performance as a function of inter-dot coupling for adiabatically driven quantum dots taking $\Gamma_M=0.05 k_{\rm B}T$. In the inset, we plot the total COP as a function of $\Gamma_M$.  All other parameters are the same as in Fig.~\ref{fig:Jiia} except $k_{\rm B}T_L=k_{\rm B}T_R =k_{\rm B}T$}
	\label{fig:cop}
\end{figure}

%\begin{figure}[!htb]
%\includegraphics[width=0.9\columnwidth]{Jiiaw.pdf}	
%	\caption{Cooling as a function of inter-dot coupling in adiabatically driven quantum dots. Parameters: $\Gamma_L=\Gamma_R=\Gamma=1$, $E_L=150\Gamma+20\Gamma\cos\left(0.1\Gamma t/\hbar\right)$, $E_R=30\Gamma+100\Gamma\cos\left(0.1\Gamma t/\hbar+\pi/2\right)$, $\Gamma_M=0.5\Gamma$, $k_{\rm B}T_L=k_{\rm B}T_R=100\Gamma$. }
%	\label{fig:Jiiaw}
%\end{figure}
In Fig.~\ref{fig:cop}, we plot the coefficient of performance as a function of the $\Gamma_M$. The red curve is the total cop whereas the black dashed curve gives the instantaneous contribution. For small values of inter-dot coupling, we observe that the cop is larger in the instantaneous case. However, in the regime where the extracted heat current shows maximum (see the bottom panel of Fig.~\ref{fig:Jiia}), the total cop (which includes both adiabatic contribution and the contribution from continuous measurement) is slightly larger than the instantaneous cop. In the inset of Fig.~\ref{fig:cop}, we plot the total cop as a function of $\Gamma_M$. The cop increases with increase in $\Gamma_M$ for small values of $\Gamma_M$ before showing a peak and then decreases monotonously. The value for which the peak is reached depends on the value of other parameters and the driving protocol considered.  

\section{Results: Coupled qubits}
In this section, we will study the effect of quantum measurement on refrigeration in coupled qubit system. Since the Hamiltonian for the coupled qubit system is similar to the coupled quantum dot system in terms of outer product operators, the behavior of extracted heat current is qualitatively similar to Figs.~\ref{fig:transient}, \ref{fig:coh} and \ref{fig:Jiia}. Similar arguments hold also for the case of coefficient of performance. For the aforementioned reasons, we will only study the effect of non-linear coupling, peculiar to bosonic systems, on extracted heat current and the co-efficient of performance.  With non-linear coupling in the right contact, the form of the contribution of the baths to the master equation in Eq.~\ref{eq:master_eq} remains intact, but the transition rates associated with the right bath changes to\cite{bibekrect}
\begin{align}
&\gamma_{{\rm R},m0}=\hbar^{-1}\lambda_{{\rm R},0m}\Gamma_{\rm R}(\epsilon_{m0}/2)\Big(1+ n_{\rm R}(\epsilon_{m0}/2)\Big)^2,\nonumber \\
&\gamma_{{\rm R},0m}=\hbar^{-1}\lambda_{{\rm R},0m}\Gamma_{\rm R}(\epsilon_{m0}/2)n_{\rm R}^2(\epsilon_{m0}/2).
\end{align}
\begin{figure}[!htb]
\includegraphics[width=\columnwidth]{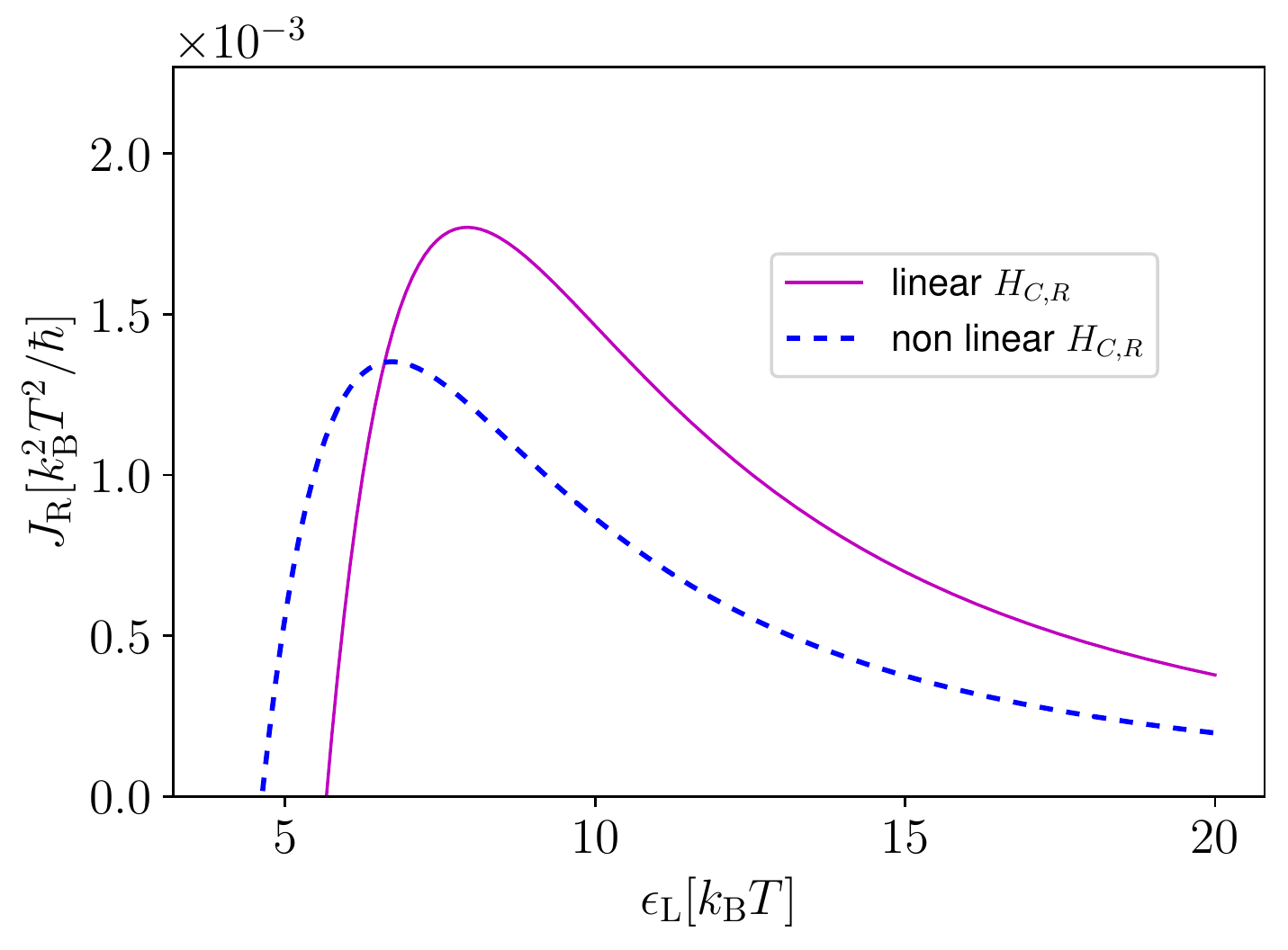}	
	\caption{Cooling as a function of $\epsilon_L$ in the coupled-qubit setup for linear and non-linear coupling betweent the system and the bath R. Parameters: $\Upsilon_L=\Upsilon_R=0.1$, $\Gamma_M=0.2 k_{\rm B}T$, $k_{\rm B}T_L=k_{\rm B}T_R=k_{\rm B}T=1$, $g=2k_{\rm B}T$, $\epsilon_R=3k_{\rm B}T$, $\epsilon_C=100k_{\rm B}T$.  }
	\label{fig:non_lin}
\end{figure}

In Fig.~\ref{fig:non_lin}, we plot the heat current flowing out of bath $R$ as a function of excitation energy of the qubit $(\epsilon_L)$ attached to the left reservoir. We consider linear coupling in the left contact whereas for the right contact we consider both linear and non-linear couplings (see Eqs.~(\ref{eq:coup_bos_lin}) and (\ref{eq:coup_bos_nonlin})). In both cases we observe that the extracted heat shows a peak for a particular value of $\epsilon_L$ before decreasing monotonously. Although the maximum with linear coupling is higher compared to the non-linear coupling case, we observe that non-linear coupling can produce cooling in the parameter regime where otherwise heating would be expected.

\begin{figure}[!htb]
\includegraphics[width=\columnwidth]{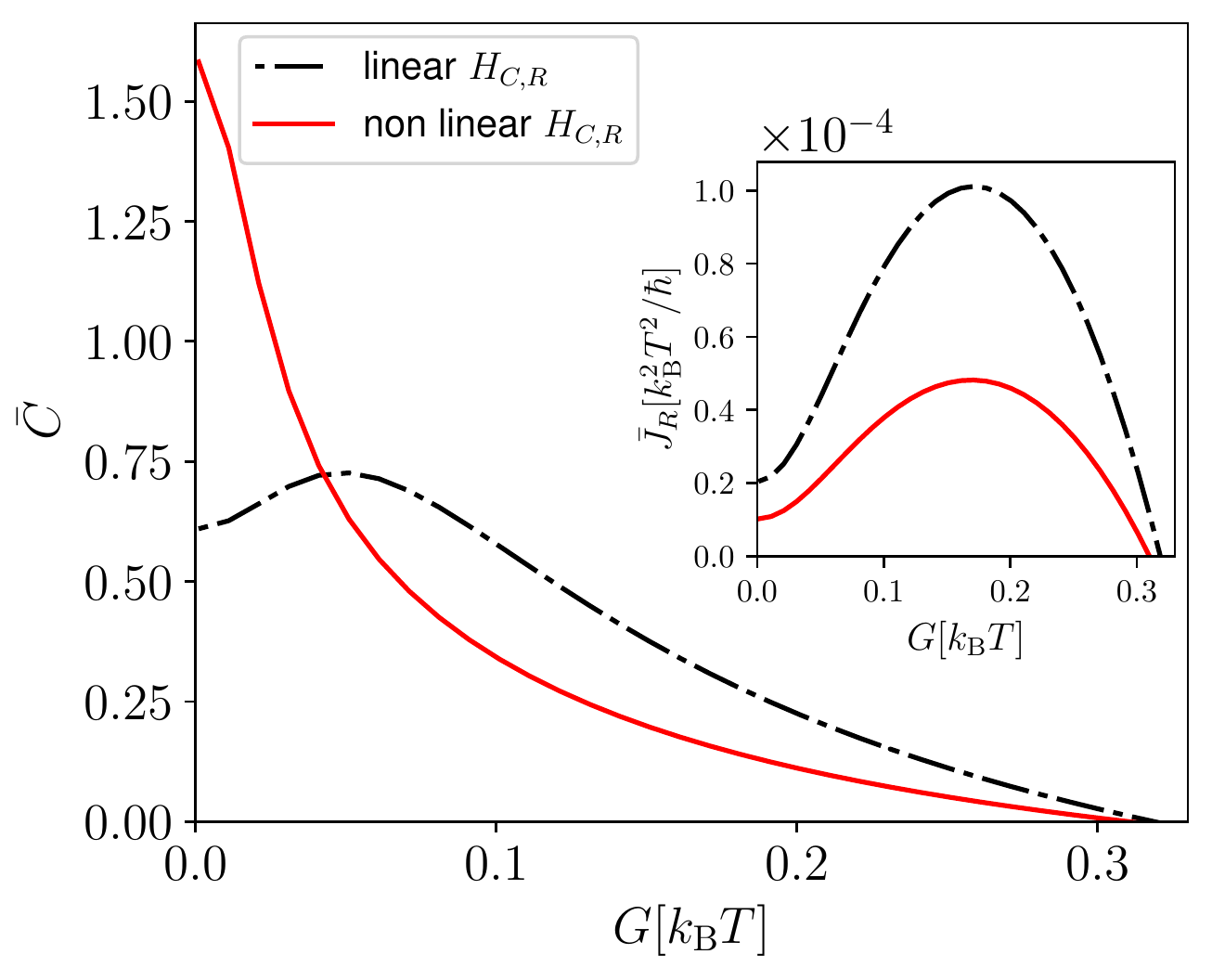}	
	\caption{Coefficient of performance as a function of inter-qubit coupling for linear and non-linear coupling betweent the coupled qubit system and the bath R. Parameters are taken similar to Fig.~\ref{fig:cop}:  $\Upsilon_L=\Upsilon_R=0.05 $, $\Gamma_M=0.05 k_{\rm B}T$, $\epsilon_L=1.5k_{\rm B}T+0.2k_{\rm B}T\cos\left(0.005k_{\rm B}T t/\hbar\right)$, $\epsilon_R=0.3k_{\rm B}T+k_{\rm B}T\cos\left(0.005k_{\rm B}T t/\hbar+\pi/2\right)$, $k_{\rm B}T_L=k_{\rm B}T_R=k_{\rm B}T$, $\epsilon_{\rm C}=10 k_{\rm B}T$.}
	\label{fig:copbos}
\end{figure}

In Fig.~\ref{fig:copbos}, we plot the coefficient of performance as a function of inter-qubit coupling strength for linear (black dashed curve) as well as non-linear (solid red curve) coupling between the coupled qubit system and the bath R. As observed in the case of coupled quantum dots, cooling is observed only for small values of $G$. For the set of parameters considered, the system goes to the heating regime for $G>0.32 k_{\rm B}T$ (see the inset for the behavior of the heat currents).  We observe that for $G<0.05 k_{\rm B}T$, non-linear coupling between the coupled qubit system and the bath $R$ gives a better coefficient of performance whereas linear coupling gives better performance in the opposite regime.  In addition, the coefficient of performance shows a peak as a function of $G$ for linear coupling. We plot the total heat current for the same set of parameters in the inset. We observe better heat extraction with linear coupling as compared to non-linear one.
\section{Conclusions}
We presented a formalism based on master equations to study the effect of continuous quantum measurement on adiabatic quantum thermal machines.  Particularly, we expressed the dynamics of density matrix in terms of the dynamics induced by the baths as well as the measurement probe for both steady state and adiabatically driven case.  We included both  average dynamics as well as the noise realizations induced by the measurement probe. In addition, using the density matrix obtained from the aforementioned quantum master equations, we defined different thermodynamics parameters, such as heat currents and coefficient of performance to describe the effect of continuous quantum measurement on adiabatic quantum refrigerators. We observed that the continuous measurement induced adiabatic heat current (discarding the measurement record) is geometric in nature. 

In order to study in detail the effect of continuous quantum measurement on energy transfer dynamics of static and driven systems, we considered two different paradigmatic examples, namely an adiabatically driven coupled quantum dot system attached to two fermionic baths and an adiabatically driven coupeld qubit system attached to two bosonic baths. In both of the examples, we studied the phenomena of refrigeration.  In contrast to absorption refrigerators where it has been observed that two linearly coupled qubits or quantum dots attached to three baths cannot produce refrigeration\cite{bibekminimal,martinez2013}, we observe that refrigeration can be achieved replacing the third bath with a measurement probe. We observe that quantum measurement alone (in the absence of adiabatic driving) can induce cooling effect in both cases of quantum dots and qubits\cite{cyrilqm}. In this paper, in addition to average dynamics we also study individual realisations when measurement records are taken into account. We observe cooling effect for a period of time in some individual realisations whereas in average heating is observed. We also study the effect of quantum coherence in the refrigeration process. We observe that quantum coherence can enhance the cooling effect. For the case of coupled qubits,we obtained cooling effect with non-linear system-bath coupling in a regime where in general (with linear coupling) heating would be observed.

In adiabatically driven systems, we observed that continuous quantum measurement can provide significant boost to the refrigeration power. Morevover, we also showed that the boost in power can be achieved without any loss in performance. Instead, for a particular set of parameters, we observed increased performance in the regime where the power of refrigeration takes a maximum.  In summary, combining two different widely used mechanisms of powering a thermal machine we obtained higher power output as well as performance. We also extended the well studied geometric aspects of adiabatic quantum thermal machines to the case where the system is continuously monitored.This provides us with a mechanism to improve heat-to-work conversion in continuosly monitored adiabatic quantum thermal machines.These findings are central to obtaining better quantum thermal machines as well as to achieving a more efficient heat management in quantum processing networks.
\begin{acknowledgments}
We thank Sreenath K. Manikandan for insightful discussions. This work was supported by the U.S. Department of Energy (DOE), Office of Science, Basic Energy Sciences (BES), under Award No. DE-SC0017890.
\end{acknowledgments}

\appendix
\begin{widetext}
\section{Calculations for quantum master equation}
\label{app:mas}

Averaging over the measurement record, the master equation consists of three terms
\begin{equation}
\frac{d\rho}{dt}=-i\left[H_S,\rho\right] + \sum_\alpha{\cal L}_\alpha\rho +{\cal L}_M\rho,
\end{equation}
where the first term on the right hand side gives the unitary dynamics within the system, the second term gives the effect of the two baths and the last term comes from the quantum measurement. For continuous measurement of the observable $X$ of the system,
\begin{equation}
{\cal L}_M\rho=\Gamma_M \left[X\rho X^\dagger-\frac{1}{2}\left(X^\dagger X\rho+\rho X^\dagger X\right)\right].
\end{equation}
Let's consider $X=|R\rangle \langle R|$, such that
\begin{align}
\langle + |{\cal L}_M\rho|+\rangle &= \Gamma_M\Big[\sin^2\theta\left(\sin^2\theta \rho_{++}+\cos^2\theta \rho_{--}+\sin\theta\cos\theta \left(\rho_{+-}+\rho_{-+}\right)\right)-\sin^2\theta \rho_{++}-\frac{1}{2}\sin\theta\cos\theta \left(\rho_{+-}+\rho_{-+}\right)\Big],\nonumber \\
\langle - |{\cal L}_M\rho|-\rangle &=\Gamma_M\Big[ \cos^2\theta\left(\sin^2\theta \rho_{++}+\cos^2\theta \rho_{--}+\sin\theta\cos\theta \left(\rho_{+-}+\rho_{-+}\right)\right)-\cos^2\theta \rho_{--}-\frac{1}{2}\sin\theta\cos\theta \left(\rho_{+-}+\rho_{-+}\right)\Big],\nonumber \\
\langle +|{\cal L}_M\rho|-\rangle &= \Gamma_M\Big[\sin\theta\cos\theta\left(\sin^2\theta \rho_{++}+\cos^2\theta \rho_{--}+\sin\theta\cos\theta \left(\rho_{+-}+\rho_{-+}\right)\right)-\frac{1}{2}\left(\sin\theta\cos\theta \left(\rho_{++}+\rho_{--}\right)+\rho_{+-}\right)\Big],\nonumber \\
\langle -|{\cal L}_M\rho|+\rangle &=\Gamma_M\Big[ \sin\theta\cos\theta\left(\sin^2\theta \rho_{++}+\cos^2\theta \rho_{--}+\sin\theta\cos\theta \left(\rho_{+-}+\rho_{-+}\right)\right)-\frac{1}{2}\left(\sin\theta\cos\theta \left(\rho_{++}+\rho_{--}\right)+\rho_{-+}\right)\Big].
\end{align}
The contribution due to the baths can be directly obtained from Ref.~\onlinecite{bibekgreen}
\begin{equation}
\langle +|{\cal L}_\alpha \rho|+\rangle=\Big(\gamma_{L,0+}+\gamma_{R,0+}\Big)\rho_{00}-\Big({\gamma}_{L,+0}+{\gamma}_{R,+0}\Big)\rho_{++}
-\frac{1}{4}\sin 2\theta \Big(-\tilde{\gamma}_{L,-0}+\tilde{\gamma}_{R,-0}\Big)(\rho_{+-}+\rho_{-+}),
\end{equation}
\begin{equation}
\langle -|{\cal L}_\alpha \rho|-\rangle=\Big(\gamma_{L,0-}+\gamma_{R,0-}\Big)\rho_{00}-\Big({\gamma}_{L,-0}+{\gamma}_{R,-0}\Big)\rho_{--}
-\frac{1}{4}\sin 2\theta \Big(-\tilde{\gamma}_{L,+0}+\tilde{\gamma}_{R,+0}\Big)(\rho_{+-}+\rho_{-+}),
\end{equation}
\begin{multline}
\langle 0|{\cal L}_\alpha \rho|0\rangle=\Big({\gamma}_{L,+0}+ \,{\gamma}_{R,+0}\Big)\rho_{++}+\Big({\gamma}_{L,-0}+{\gamma}_{R,-0}\Big)\rho_{--}-\Big({\gamma}_{L,0+}+{\gamma}_{R,0+}+{\gamma}_{L,0-}+{\gamma}_{R,0-}\Big)\rho_{00}\\
-\frac{1}{4}\sin 2\theta \Big(-\tilde{\gamma}_{L,-0}-\tilde{\gamma}_{L,+0}+\tilde{\gamma}_{R,-0}+\tilde{\gamma}_{R,+0}\Big)(\rho_{+-}+\rho_{-+}),
\end{multline}
and
\begin{multline}
\langle +|{\cal L}_\alpha \rho|-\rangle=\frac{i}{\hbar}\epsilon_{+-}\rho_{+-}+\frac{1}{4}\sin2\theta\bigg[\Big(-\tilde{\gamma}_{L,0-}-\tilde{\gamma}_{L,0+}+\tilde{\gamma}_{R,0-}+\tilde{\gamma}_{R,0+}\Big)\rho_{00}
+\Big(\tilde{\gamma}_{L,+0}-\tilde{\gamma}_{R,+0}\Big)\rho_{++}+\Big(\tilde{\gamma}_{L,-0}-\tilde{\gamma}_{R,-0}\Big)\rho_{--}\bigg]\\
-\frac{1}{2}\Big({\gamma}_{L,-0}+{\gamma}_{L,+0}+{\gamma}_{R,-0}+{\gamma}_{R,+0}\Big)\rho_{+-},
\end{multline}
where 
\begin{align}
&\tilde{\gamma}_{\alpha,m0}=\hbar^{-1}\Gamma_\alpha(\epsilon_{m0})\Big(1\pm n_\alpha(\epsilon_{m0})\Big),\nonumber \\
&\tilde{\gamma}_{\alpha,0m}=\hbar^{-1}\Gamma_\alpha(\epsilon_{m0})n_\alpha(\epsilon_{m0}).
\end{align}
In the steady state, $\frac{d\rho}{dt}=0$. Using this relation, one can write the off-diagonal terms of density matrix in terms of the diagonal terms.  Once the resultant expressions for off-diagonal terms of density matrices are substituted into the master equations for the population (diagonal terms), the master equations can be readily solved\cite{bibekgreen}.

In a similar manner, the heat current flowing into the bath $R$ can be expressed as
\begin{equation}
J_R=\epsilon_{+0}\theta\Big( {\gamma}_{R,+0}\rho_{++}-{\gamma}_{R,0+}\rho_{00}\Big)+\epsilon_{-0}\Big( {\gamma}_{R,-0}\rho_{--}-{\gamma}_{R,0-}\rho_{00}\Big)
+\frac{1}{4}\sin 2\theta\Big(\epsilon_{+0}\tilde{\gamma}_{R,+0}+\epsilon_{-0}\tilde{\gamma}_{R,-0}\Big)\left(\rho_{+-}+\rho_{-+}\right)
\end{equation}
\end{widetext}

\bibliography{paper}

%\begin{thebibliography}{9}
%\bibitem{bibekgrn}
%B. Bhandari, R. Fazio, F. Taddei and L. Arrachea, Phys. Rev. B, {\bf 104}, 035425 (2021)
%
%\bibitem{bibekgeo}
%B. Bhandari, P. T. Alonso, F. Taddei, F. v. Oppen and R. Fazio, Phys. Rev. B, {\bf 102}, 155407 (2020)
%
%\bibitem{breuer}
%H. P. Breuer and F. Petruccione, The theory of open quantum systems, Oxford University Press on Demand (2002)
%
%\bibitem{jacobs}
%K. Jacobs,Quantum measurement theory and its applications, Cambridge University Press (2014)
%\end{thebibliography}

\end{document}